\documentclass[
 reprint,
 superscriptaddress,
 amsmath,amssymb,
 aps,
 prb,
]{revtex4-2}

\usepackage{tikz}
\usetikzlibrary{shapes.geometric}
\usetikzlibrary{backgrounds}
\tikzset{
  tensor network/.style={
    every node/.style={circle, draw, minimum size=0.6cm},
    edge/.style={thick},
    triangle_r/.style={
        draw,
        fill=blue!20,
        regular polygon, 
        regular polygon sides=3, 
        minimum size=0.8cm, 
        inner sep=0pt,
        shape border rotate=90,
    },
    triangle_l/.style={
        draw,
        fill=blue!20,
        regular polygon, 
        regular polygon sides=3, 
        minimum size=0.8cm, 
        inner sep=0pt,
        shape border rotate=270,
    }
  }
}

\usepackage{braket}
\usepackage{float}
\usepackage{algorithm}
\usepackage{algpseudocode}

\usepackage{physics}
\usepackage{subfig}
\usepackage{ulem}
\renewcommand{\emph}[1]{{\it #1}}
\usepackage[english]{babel}      
\makeatletter
\@ifundefined{l@en}{\let\l@en\l@english}{}
\@ifundefined{captionsen}{}{}
\@ifundefined{dateen}{}{}
\@ifundefined{extrasen}{}{}
\@ifundefined{noextrasen}{}{}
\makeatother
\usepackage{graphicx}
\usepackage{dcolumn}
\usepackage{bm}

\usepackage[colorlinks=true,
            citecolor=blue,
            linkcolor=blue,
            urlcolor=blue,
            filecolor=blue,
            pdfborder={0 0 0}]{hyperref}
\makeatletter
\long\def\@makecaption#1#2{%
  \par
  \vskip\abovecaptionskip
  \begingroup
    \small\rmfamily
    \samepage
    \flushing
    \let\footnote\@footnotemark@gobble
    \@make@capt@title{#1}{#2}\par
  \endgroup
  \vskip\belowcaptionskip
}
\makeatother
\begin{document}

\newcommand{\TZ}[1]{{\color{blue}#1}}

\title{\textbf{Critical behaviors of magic and participation entropy at measurement induced phase transitions}}%
\author{Eliot Heinrich}
\affiliation{Department of Physics, Boston College, Chestnut Hill, MA 02467, USA}
\author{Hanchen Liu}
\affiliation{Department of Physics, Boston College, Chestnut Hill, MA 02467, USA}
\author{Tianci Zhou}
\affiliation{Department of Physics, Virginia Tech, Blacksburg, Virginia 24061, USA}
\author{Xiao Chen}
\affiliation{Department of Physics, Boston College, Chestnut Hill, MA 02467, USA}

\date{\today}
\begin{abstract}

We study the participation and stabilizer entropy of non-unitary quantum circuit dynamics, focusing on the critical line that separates the low-entanglement spin-glass phase and the paramagnetic phase. Along this critical line, the entanglement has a logarithmic scaling, which enables us to access the critical regime using large-scale matrix product state simulations with modest bond dimension. We find that both the participation entropy and stabilizer entropy exhibit critical slowing down: their saturation time scales linearly with the system size, in stark contrast to purely unitary dynamics, where saturation occurs on logarithmic time scales. In addition, we study bipartite participation and stabilizer mutual information, and find that it shows similar scaling behavior to the entanglement entropy. Finally, by analyzing the participation entropy of several paradigmatic Clifford circuits, we identify similar slow dynamical behavior near their respective critical points.
\end{abstract}

\maketitle


	 


\section{Introduction}

Magic quantifies the non-stabilizerness of quantum states and is a resource for quantum computation~\cite{gu_magic_2024, oliviera_magic_2022, bravyi_magic_2005, szombathy2025}. Highly entangled states are generally difficult to simulate classically, yet there exist a large class of exceptions, called stabilizer states~\cite{aaronson_2004}. These states may possess volume-law entanglement, yet their evolution under Clifford circuits remains as stabilizer states. Those states can be efficiently tracked using linear algebra over Pauli operators, requiring only polynomial resources in the number of qubits~\cite{nahum_entanglement_2017, gutschow_entanglement_2010}. Consequently, universal quantum computation has to extend beyond stabilizer states. And the ``distance'' of a state from the stabilizer polytope provides a notion of classical simulation hardness. This defines the computational aspect of magic.

Several measures have been proposed to quantify magic, including mana, robustness of magic, and various distances to the stabilizer polytope~\cite{liu_magic_2022, wang_magic-2019, hamaguchi_magic_2024, haug_magic_2023}. In this work, we focus on the \emph{stabilizer R\'enyi entropy} (SRE), a measure which is physically intuitive and numerically convenient to compute~\cite{lorenzo_sre_2022, ding_sremc_2025, tarabunga_magic_2025, lipardi2025, liu2025_neq_sre_mc}. Taking a pure state of an $N$-qubit system, its density matrix can always be expanded in the Pauli basis as
\begin{equation}
    \rho = \frac{1}{2^{\frac{N}{2}}}\sum_{P} c_P P,
\end{equation}
where the sum runs over all $4^N$ Pauli strings. Interpreting $|c_P|^2$ as a probability distribution, the SRE is defined as the R\'enyi entropy of this distribution minus $N\ln 2$. Stabilizer states correspond to an equal-weight superposition of $2^N$ Pauli strings and thus the subtraction yields zero SRE, whereas Haar-random states spread nearly uniformly over all $4^N$ strings and give an SRE $\approx N \ln 2$. Therefore, SRE provides an operational measure of a state’s distance from the stabilizer states and quantifies its classical intractability.

Recent studies have revealed that the dynamics of magic can differ dramatically from that of entanglement. In local random unitary circuits, entanglement grows linearly in time and saturates on a timescale $t \sim N$~\cite{keyserlingk2018_ballistic, piroli2020_ballistic, chan2018_ballistic, kim2013_ballistic}. In contrast, even one layer of single-qubit non-Clifford gates can generate extensive magic. In generic random circuits its relaxation to the steady-state value occurs on a much shorter timescale $t \sim \ln N$~\cite{turkeshi_magic_2025}. This rapid generation and relaxation indicate that magic probes different aspects of quantum complexity compared to entanglement.

{Motivated by these observations, we investigate the dynamics of magic and compare it with entanglement at a \emph{measurement-induced phase transition} (MIPT).} It is well known that in hybrid unitary-projective circuits, tuning the measurement rate drives a transition between a volume-law entangled phase and an area-law phase~\cite{skinner_mipt_2019, iaconis_mipt_2020, bao_mipt_2020, choi_mipt_2020, koh_mipt_2023}. How does magic behave in non-unitary dynamics? Specifically, how does magic scale near these critical points, and can it serve an independent probe of MIPT criticality? 

To address these questions, we study the dynamics of SRE in a hybrid circuit with a Kramers-Wannier duality symmetry. In addition to volume-to-area law transition, the model also has an area-to-area law transition regime that allows simulations of large system sizes using matrix-product-state (MPS) techniques. The weak measurements in the circuit always generate volume law magic, which is different from the purification setups in Refs.~\cite{paviglianiti_estimating_2024,bejan_dynamical_2024} where projective Pauli measurements only suppress magic and an area-law magic phase is possible. By a perfect-sampling method~\cite{lami2023}, we numerically compute the SRE and demonstrate that magic exhibits \emph{critical slowing down} near the transition, in sharp contrast to its fast relaxation in generic unitary circuits. We extract the corresponding scaling behaviors and dynamical exponents. Additionally, we show that the bipartite mutual information of SRE [termed the \emph{stabilizer mutual information} (SMI)~\cite{ding_sremc_2025}] exhibits critical scaling similar to entanglement entropy. Thus the non-local portion of magic captured by SMI detects the MIPT criticality. Together, these results demonstrate that  magic serves as an independent diagnostic of non-equilibrium quantum criticality in hybrid circuits.

Alongside SRE, we also study the \emph{participation entropy} (PE), which characterizes the spread of the wavefunction in the computational basis. For a pure state $|\psi\rangle = \sum_z \psi_z |z\rangle$, the PE is defined as the R\'enyi entropy of the probability distribution $|\psi_z|^2$. PE has been widely used to study localization and anti-concentration properties of the measurement probabilities~\cite{beugeling_pe_2015, liu_pe_2025}. Such anti-concentration is typically viewed as a signature of complexity in the unitary evolution. It manifests in the distribution of the measurement outcomes of a random circuit sampling and has been proven achievable in a logarithmic circuit depth~\cite{dalzell_random_2022-1,Magni_2025}. More recent results show that anti-concentration implies a state 2-design~\cite{heinrich_anticonc_2025}. The logarithmic depth indicates that PE grows rapidly in generic random circuits and saturates on a timescale $t \sim \ln N$, similar to the dynamics of SRE, yet in contrast to the linear entanglement growth. This difference arises because both PE and SRE can be generated locally by basis mixing, whereas entanglement requires gates that act across a bipartition. 

From the definition, both SRE and PE are entropy measures of mixing in the quantum systems: the former in the operator Pauli basis and the latter in the state computational space. Thus, PE is a simpler proxy to understand the dynamics of both. One observation for PE, which is also its distinction to the SRE, is that it remains nonzero for stabilizer states. This offers a way to compute PE  efficiently for large systems in the stabilizer formalism. Motivated by this feature, we investigate the dynamics of PE at the MIPT critical points in various hybrid Clifford circuits, with the expectation that these results provide insights into both magic and PE in more generic non-Clifford dynamics. In particular, we analyze critical points of hybrid random Clifford circuits with dynamical exponent $z=1$ and  $z>1$. We also collect the data of  PE using the perfect-sampling method~\cite{tarabunga_many-body_2023}  along with the SRE in non-Clifford hybrid circuits with the duality symmetry in the previous paragraph. 

Across all these models, we find that PE exhibits similar critical scaling behaviors to SRE. We observe that although PE rapidly grows to a volume-law value within $\mathcal{O}(1)$ layers, its relaxation toward saturation at the critical point is governed by a universal scaling function of $L/t^z$. Furthermore, we introduce the \emph{participation mutual information} (PMI) and show that for the Clifford circuits, at criticality, the bipartite PMI grows logarithmically in time, mirroring the behavior of SMI and entanglement.



\section{Definition of entropies}

In this work, entanglement, magic and anti-concentrations of the measurement outcome distributions are all measured through 
a R\'enyi entropy over relevant distributions. The $n$-R\'enyi entropy of a distribution ${q : \mathcal{X} \rightarrow \mathbb{R}[0,1]}$ is defined as
\begin{align}\label{eq:renyi_entropy}
    S_n(q) = 
    \begin{cases}
            -\sum\limits_{x \in \mathcal{X}} q(x) \log q(x) & n = 1 \\
        \frac{1}{1 - n}\log\sum\limits_{x \in \mathcal{X}} q(x)^n & n \neq 1
    \end{cases}
\end{align}
We use 
R\'enyi entanglement entropy (EE) to characterize entanglement. 
A pure state $|\psi\rangle$ always has a Schmidt decomposition with respect to a bipartition into subsystems $A$ and $B$
\begin{align}\label{eq:schmidt}
    \ket{\psi} = \sum\limits_{i=1}^m \lambda_i \ket{u_i}_A\ket{v_i}_B
\end{align}
where $\ket{u_i}_A$ and $\ket{v_i}_B$ are orthonormal states in $A$ and $B$, respectively. By conservation of total probability, the singular values $\lambda = (\lambda_1, ..., \lambda_m)$ squared form a normalized distribution, i.e. $\sum\limits_i \lambda_i^2 = 1$. The R\'enyi entanglement entropy of ${\rho}$ with respect to the bipartition formed by $A$ and $B$ is
\begin{align}\label{eq:ee}
    S_n^{\mathrm{EE}}(\rho) = S_n(\lambda^2)
\end{align}

We use Stabilizer R\'enyi Entropy (SRE)~\cite{lorenzo_sre_2022} to characterize the non-stabilizerness of a quantum state, the magic. The SRE is defined, for an $L$-qubit quantum state, as
\begin{align}\label{eq:sre}
    S^{\mathrm{SRE}}_n(\rho) = -L\log2 + S_n(\Pi_\rho)
\end{align}
Here $\rho$ can be a mixed state and $\Pi_{\rho}$ is the distribution of $\rho$  over $L$ -qubit Pauli strings  ${\boldsymbol\sigma = \sigma_1\cdots\sigma_L}$ 
\begin{equation}
    {\Pi_\rho(\boldsymbol \sigma) = \frac{1}{2^L}|\text{Tr}(\rho \boldsymbol \sigma)|^2}/\mathrm{Tr}(\rho^2).
\end{equation}
For pure states, $\mathrm{Tr}(\rho^2) = 1$

To distinguish the local and non-local magic, we introduce the stabilizer mutual information (SMI):
\begin{align}\label{eq:mutual_sre}
    L^{\mathrm{SRE}}(\rho_{AB}) = S_2^{\mathrm{SRE}}(\rho_A) + S_2^{\mathrm{SRE}}(\rho_B) - S_2^{\mathrm{SRE}}(\rho_{AB})
\end{align}
In this paper, we focus on the bipartite SMI (BSMI), i.e. $B = \bar{A}$, the complement of $A$.

Finally, we study the Participation Entropy (PE), defined as
\begin{align}\label{eq:pe}
    S^{\mathrm{PE}}_n(\rho) = S_n(p), \quad p(z) = \mathrm{Tr}(\rho\ket{z}\bra{z})
\end{align}
where $z$ is an $L$-bit bitstring: 
\begin{align}
    \ket{z} = \bigotimes\limits_{i=1}^L \ket{z_i}, \quad Z_i \ket{z_i} = z_i\ket{z_i}, \quad z_i=\pm 1.
\end{align}

Similarly to the SMI, we define the participation mutual information (PMI):
\begin{align}\label{eq:mutual_pe}
    L^{\mathrm{PE}}(\rho_{AB}) = S^{\mathrm{PE}}_1(\rho_A) + S^{\mathrm{PE}}_1(\rho_B) - S^{\mathrm{PE}}_1(\rho_{AB}).
\end{align}
Again, we focus on the bipartite PMI (BPMI) of a quantum state with $B = \bar{A}$.

\section{Self-dual hybrid circuit}
\label{sec:self_dual}

\begin{figure}[h!]
  \centering
  \includegraphics[width=0.8\linewidth]{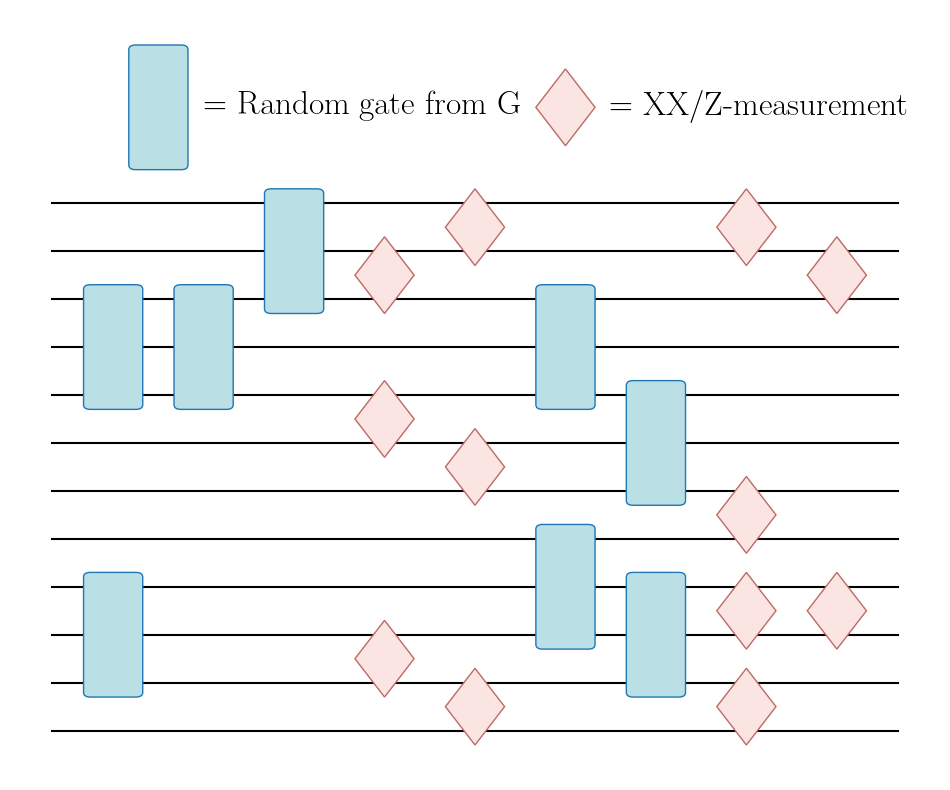}
  \caption{An example of the hybrid circuit dynamics described in the text. Time progresses from left to right. The blue rectangles correspond to Clifford unitaries selected randomly from the self-dual ensemble described in the text. The red rhombuses correspond to weak measurements with strength $\beta$ in the basis of either $XX$ or $ZI$, with probability $p$ and $1 - p$, respectively. }
  \label{fig:z2_model}
\end{figure}

\begin{figure}[h!]
  \centering
  \includegraphics[width=\linewidth]{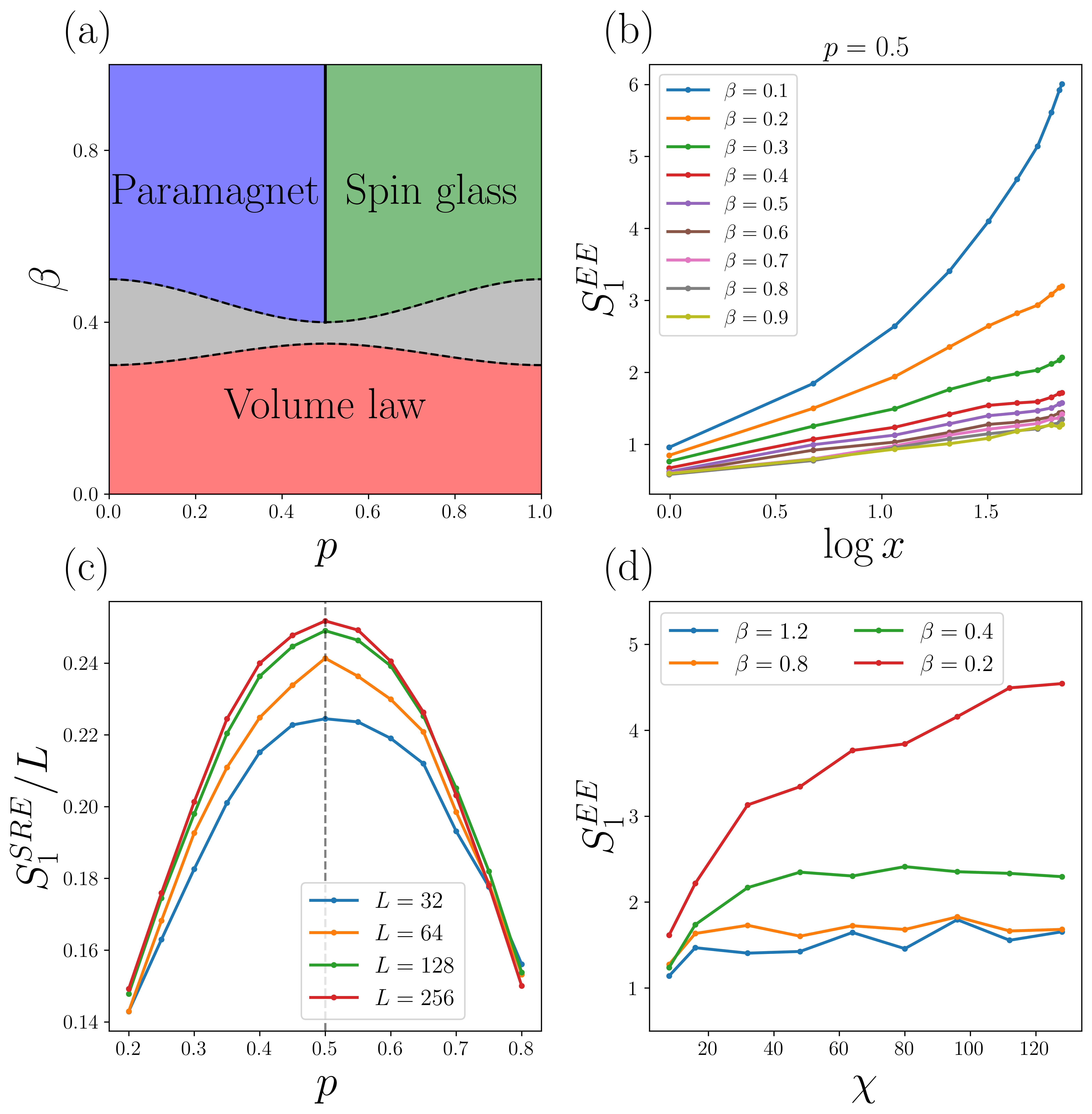}
  \caption{ (a) A schematic picture of the phase diagram. The model with $Z_2$ symmetry exhibits three phases: a paradigmatic volume-law phase (in red), a paramagnetic area-law phase (in blue), and a spin-glass area-law phase (in green). The transition between the paramagnetic and spin-glass phase occurs at $p = 0.5$. The gray region separating the volume-law from the area-law phases indicates uncertainty in $\beta_c$. (b) We show the entanglement as a function of $\log x$ for a small system of $L=24$ qubits and $p = 0.5$. The results are computed by equilibrating the system for $t_{eq} = 100$ steps, and then sampling the entanglement every $t_m = 5$ steps for an additional $t_s = 100$ steps. (c) We show the SRE density at $\beta = 0.8$ for various system sizes and as $p$ is varied. We see that the SRE is extensive for all $p$. We set $t_{eq} = 500$, $t_m = 25$, $t_s = 500$. (d) We show the converged entanglement entropy as a function of the max bond dimension imposed on the system for $p = 0.5$ and $L = 128$. We set $t_{eq} = 500$, $t_m = 25$, $t_s = 500$.}
  \label{fig:phase_diagram}
\end{figure}

\begin{figure}[h!]
  \centering
  \includegraphics[width=\linewidth]{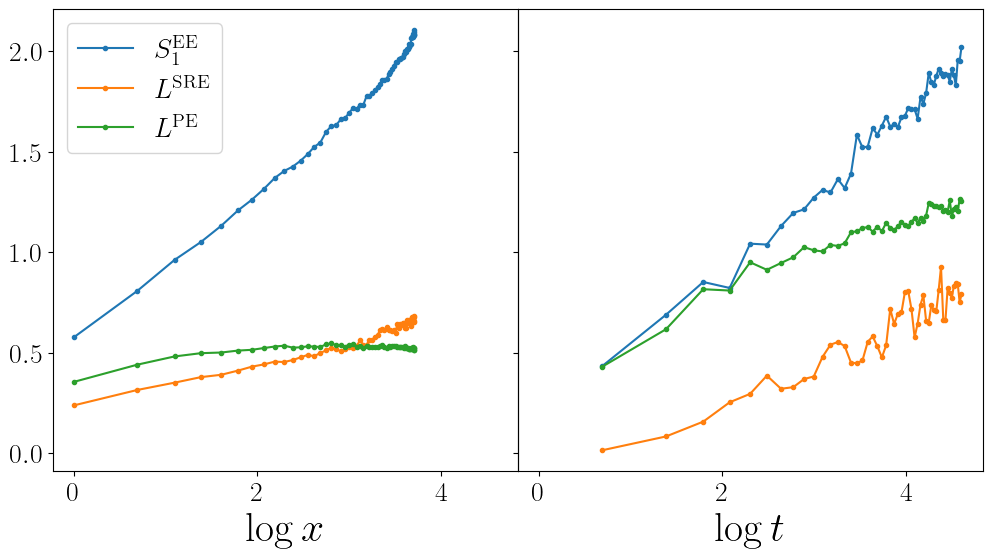}
  \caption{Growth of various entropies and bipartite mutual information as functions of time and subsystem size for a system of size $L = 128$ at $p = 0.5$ and $\beta = 0.8$. (Left) Steady-state scaling of the von Neumann EE, BSMI, and BPMI as the bipartition size $\ell$ is varied. (Right) The same quantities exhibit logarithmic growth in time. The scaling coefficients of the EE are $\alpha_t^{\mathrm{EE}} = 0.41$ and $\alpha_s^{\mathrm{EE}}=0.42$, yielding $z^{\mathrm{EE}} = 1.02$. The scaling coefficients of the SMI are $\alpha_t^{\mathrm{SRE}} = 0.22$ and $\alpha_s^{\mathrm{SRE}} = 0.18$, yielding $z^{\mathrm{SRE}} = 0.81$. The deviation from logarithmic behavior observed in the PE is discussed in the main text. }
  \label{fig:scaling}
\end{figure}

\begin{figure}[h!]
  \centering
  \includegraphics[width=\linewidth]{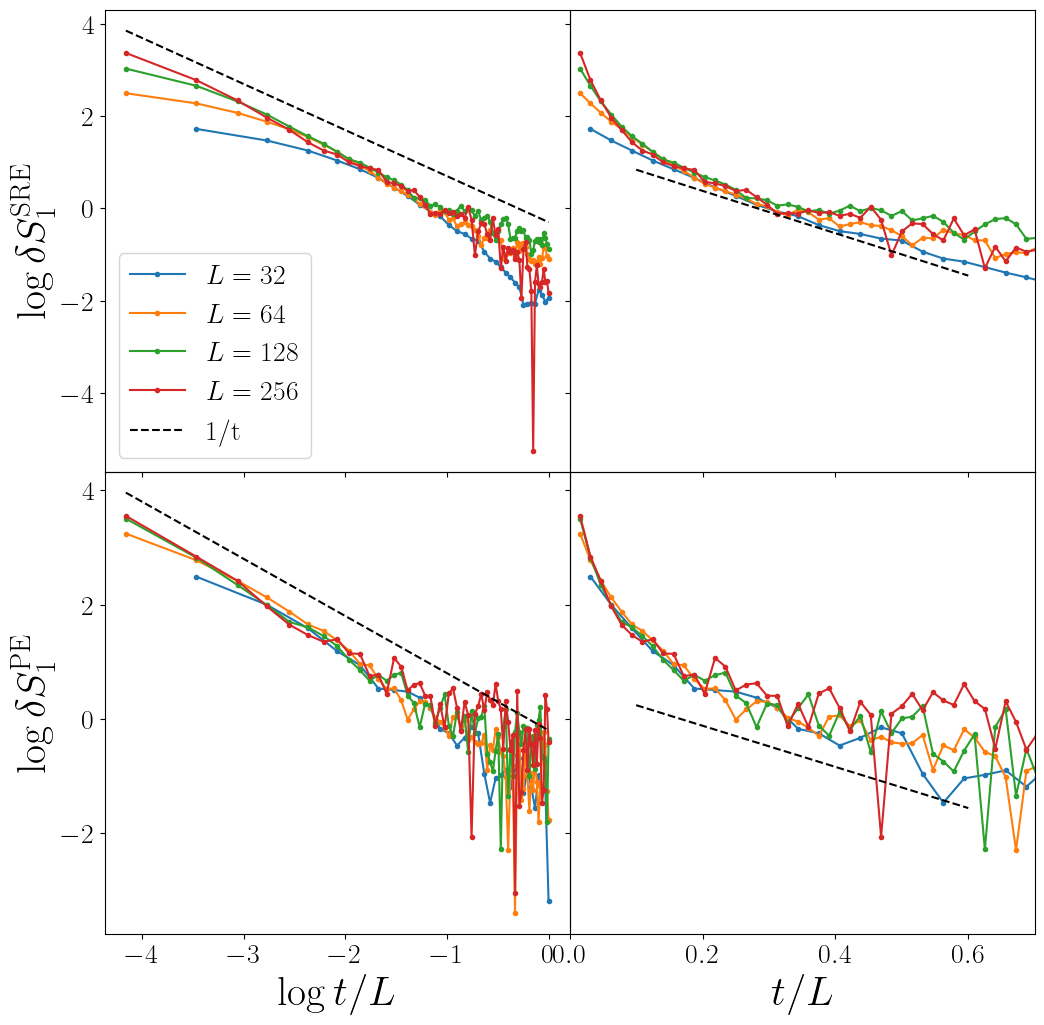}
  \caption{We show the approach to equilibrium by the SRE and PE. The measurement strength parameter $\beta = 0.8$. For each reported value, the entropies are calculated according to the methods described in Apps.~\ref{app:stabilizer_entropy}-\ref{app:participation_entropy}. The left column shows the data on a log-log plot, emphasizing that at early times, the deviation from the steady state scales according to a power-law, while the right column is on a semi-log plot, emphasizing the crossover to an exponential approach to equilibrium at later times. In both columns, we have added dotted lines to serve as a guide to linear behavior indicating power-law (exponential) behavior on the left (right) column. On the left, the slope is fixed at 1 to emphasize the conformal symmetry. }
  \label{fig:equilibrium}
\end{figure}

In this section, we consider a 1+1D random Clifford circuit interspersed with weak measurement gates. As illustrated in Fig.~\ref{fig:z2_model}, the circuit consists of alternating rounds of unitary gates and weak measurements denoted as cyan and pink boxes respectively. Each unitary round has  $L/3$  three-qubit gates and each measurement round has $L/2$ two-qubit weak measurements. All the gates are applied in random positions. 

Precisely locating the critical point of an MIPT is a necessary first step for observing its critical behavior, but doing so adds extra difficulty to the already computationally intensive task of evaluating the SRE. Therefore we choose the ensemble of unitaries and weak measurements to have a duality symmetry which fixes the critical point exactly. Inspired by Ref.~\cite{sang_measurement_protected_2021}, the model is chosen to have a Kramers-Wannier duality symmetry $Z_i \longleftrightarrow X_i X_{i+1}$, which corresponds to a shift of one lattice site of the Majoranas in the Jordan-Wigner transformation of the circuit dynamics; see App.~\ref{app:majorana} for details. In the spin basis, the duality transform maps
\begin{align*}
    Z_i &\rightarrow X_i X_{i+1}, \\
    Z_i Z_{i+1} &\rightarrow X_i X_{i+2} 
\end{align*}
We therefore choose our unitaries to be from an ensemble of Clifford gates 

\begin{align}
    S_i = \{e^{i \frac{\pi}{4} Z_i}, e^{i \frac{\pi}{4} X_i X_{i+1}}, e^{i \frac{\pi}{4} Z_i Z_{i+1}}, e^{i \frac{\pi}{4} X_i X_{i+2}}\}
\end{align}
and the corresponding hermitian conjugates $S^\dagger_i = \{  U \in S_i\ : U^\dagger\}$. The full gate set is then $G = S_i \cup S_i^\dagger$.

To fit the three-qubit boxes in Fig.~\ref{fig:z2_model}, identity operators are appended to the right when necessary.  For example, $e^{i \frac{\pi}{4} Z_i}$ is understood as $e^{i \frac{\pi}{4} Z_i} \otimes I_{i+1} \otimes I_{i+2}$ in the diagram.  Gates are chosen from this ensemble with equal probability  so that every gate and its dual is equally likely to be applied. In the open boundary condition we choose, the unitary dynamics are overall self-dual. 

The weak measurements are specified by Kraus operators 
\begin{align}
    K_{P,+} = \frac{e^{\beta P}}{\sqrt{2\cosh(2\beta)}}, \quad K_{P,-} = \frac{e^{-\beta P}}{\sqrt{2 \cosh(2\beta)}},
\end{align}
where $P$ is a two-qubit Pauli operator. With probability $p$, we take $P = Z_i \otimes I_{i+1}$, and with probability $1 - p$, we take $P = X_i X_{i+1}$. 

We simulate the pure-state trajectories by randomly applying $K_{P,+}$ or $K_{P,-}$ to the state $\ket{\psi}$ with probability 
\begin{align}
    q_{P,\pm} = \bra{\psi} K_{P,\pm}^2\ket{\psi} = \frac{1}{2}\left[1 \pm \tanh(2\beta) \bra{\psi} P \ket{\psi}\right]
\end{align}
followed by re-normalization of the state. In practice, we map
\begin{align}
    \ket{\psi} \rightarrow \ket{\tilde{\psi}_{P,\pm}} = \frac{e^{\pm \beta P} \ket{\psi}}{\sqrt{\bra{\psi} e^{\pm 2 \beta P}\ket{\psi}}}
\end{align}
with probability $q_{P,\pm}$. In the limit of the measurement strength $\beta \rightarrow \infty$, these measurements become projective. This layer of measurements is evidently self-dual at $p = 0.5$ under the transformation $Z_i \longleftrightarrow X_i X_{i+1}$. Since the unitary dynamics are self-dual, the circuit dynamics as a whole are therefore self-dual at $p = 0.5$.

This model is a variation on the pure Clifford one studied by Ref.~\cite{sang_measurement_protected_2021}; the primary differences are the more restrictive set of unitaries and the non-Clifford measurement operations. Nonetheless, it exhibits a similar phase diagram with  three phases. As shown in Fig.~\ref{fig:phase_diagram}(a), increasing the measurement strength $\beta$ leads to an entanglement transition from the volume-law entangled phase to the area-law disentangled phase at $\beta_c(p)$. Above $\beta_c(p)$, the model exhibits an area-law to area-law phase transition between a $Z$-measurement dominated paramagnetic phase and an $XX$-measurement dominated spin-glass phase as $p$ is varied. The phase transition separating the two area-law phases occurs precisely at $p = 0.5$ due to the imposed duality symmetry. This features a key advantage of the model, as it eliminates the need for numerically locating the critical point. In contrast, the precise location of the transition $\beta_c(p)$ between the area-law and volume-law phases is harder to determine numerically. 

A schematic phase diagram is shown in Fig.~\ref{fig:phase_diagram}(a). In Fig.~\ref{fig:phase_diagram}(b), we estimate that $\beta_c(0.5) \sim 0.4$, corresponding to the tricritical point, by observing that the entanglement grows faster than logarithmically when $\beta \lesssim 0.4$. That is, for $\beta \lesssim 0.4$ and $p = 0.5$, the system is in a paradigmatic volume law phase, and for $\beta \gtrsim 0.4$, a critical line lying between the spin-glass and paramagnetic phases. Different from the models studied by Ref.~\cite{sang_measurement_protected_2021,paviglianiti_estimating_2024,bejan_dynamical_2024}, the weak measurements in our circuit introduce non-stabilizerness into the system, and stabilizer simulations are not feasible. In fact, the magic is extensive throughout the area-law phases, as can be seen in Fig.~\ref{fig:phase_diagram}(c). We focus on the area-law phases and the critical line between them, where entanglement is low and we can perform large-scale numerical simulations using MPS techniques with modest bond dimension. We take open boundary conditions in the simulation. In Fig.~\ref{fig:phase_diagram}(d), we observe that for $\beta \gtrsim 0.4$, a maximum bond dimension of $\chi > 32$ is sufficient for numerical convergence of the entanglement entropy. Thus, we fix $\chi = 128$ for the remainder of the work.

We first compute the entanglement entropy $S^{\mathrm{EE}}_1(\ell)$ as defined in Eq.~\eqref{eq:ee}. We consider subsystem $A$ to be the leftmost $\ell$ qubits and its complement $\bar{A}$ to be the rightmost $L - \ell$ qubits. Along the critical line separating the two area-law phases, we find that ${S^{\mathrm{EE}}_1(L/2) = \alpha_t \log t}$ for early times, and saturates to a scaling form ${S^\mathrm{EE}_1(\ell) =\alpha_s \log x}$, where 
\begin{align}
\label{eq:x_scaling}
    x = \frac{L}{\pi} \sin\left(\frac{\ell \pi}{L}\right).
\end{align} 
We find that $z = \alpha_s/\alpha_t \sim 1$, suggesting the critical line has conformal symmetry. The result for $\beta=0.8$ is shown in Fig.~\ref{fig:scaling}. Below, we focus on this point and analyze its SRE and PE.

We compute the total SRE for the wave function $\ket{\psi(t)}$. The SRE from Eq.~\ref{eq:sre} can be estimated as 

\begin{align}\label{eq:sre_est}
    S^{\mathrm{SRE}}_n(\rho) = -L\log2 +  
    \begin{cases}
        -\left\langle \log \Pi_\rho\right\rangle_{\Pi_\rho} & \text{if } n = 1,\\
        \frac{1}{1 - n} \log\left\langle \Pi_\rho^{n - 1}\right\rangle_{\Pi_\rho} & \text{if } n \neq 1
    \end{cases}
\end{align}
by drawing samples of Pauli operators $\boldsymbol \sigma$ from the distribution $\Pi_\rho(\boldsymbol \sigma)$. This can be accomplished for an MPS simulation using the perfect-sampling algorithm of Ref.~\cite{lami2023}, which is summarized in App.~\ref{app:stabilizer_entropy}. The core of the algorithm samples Pauli strings by sequentially computing the conditional probability of the Pauli on each site through the chain rule of a joint distribution. We use $N_s = 5000$ Pauli string samples for the stabilizer entropy for all samples in this text. 

Due to weak measurements which can locally generate magic, SRE becomes extensive even in the area-law phases. However, unlike in the case of random Haar unitary circuit and Floquet dynamics where the SRE approaches its steady-state value in just $\log L$ time~\cite{tirrito2025, turkeshi_magic_2025}, our non-unitary dynamics exhibit a much slower relaxation to equilibrium. To quantify this behavior, we define and compute
\begin{align}
    \delta S^{\mathrm{SRE}}_1(t) \equiv |S^{\mathrm{SRE}}_1(t)-S^{\mathrm{SRE}}_1(\infty)|.   
\end{align}

We find that the relaxation to equilibrium is significantly slower, as illustrated in Fig.~\ref{fig:equilibrium}. The deviation $\delta S^{\mathrm{SRE}}_1$ collapses onto a universal scaling function of a {\it single} variable $\tau=t/L$. In the early-time regime ($\tau\ll 1$), we observe a clear power law decay 
\begin{align}
    \delta S^{\mathrm{SRE}}_1(\tau) \sim 1/\tau.
\end{align} 
At later times ($\tau \gg 1$), the observed decay approximately crosses over to an exponential form, i.e.,
\begin{align}
    \delta S^{\mathrm{SRE}}_1(\tau) \sim\exp(-a\tau).
\end{align} 
A similar behavior is found for the dynamics of PE, which we estimate using the algorithm in App.~\ref{app:participation_entropy} using $N_s = 5000$ bitstring samples. The deviation from equilibrium for PE, 
\begin{align}
\delta S^{\mathrm{PE}}_1(t)\equiv|S^{\mathrm{PE}}_1(t) - S^{\mathrm{PE}}_1(\infty)|    
\end{align}
also collapses as a function of $\tau=t/L$, showing $1/\tau$ decay at early times and an exponential tail at later times.

These results indicate that the critical point exhibits critical slowing down in both SRE and PE, with each requiring $O(L)$ time to saturate to their respective steady-state values. We note that the exponential decay tail exhibits noticeable fluctuations due to its small magnitude, making it difficult to resolve using our sampling method. This tail becomes more apparent in the subsequent analysis in a purely Clifford circuit (including the measurements), where PE can be computed exactly.

Both the SRE and PE are extensive in the area-law phases, but they are largely localized and there is a delocalization transition across the critical point. This structural change can be characterized by the BSMI as defined in Eq.~\eqref{eq:mutual_sre}, where local contributions are subtracted. Similar to the entanglement, the BSMI and BPMI are close to zero in the paramagnetic phase, and saturate to a finite constant in the spin glass phase. As shown in Fig.~\ref{fig:scaling}, at the critical point, the BSMI grows logarithmically with time before saturating. For the steady state, we find that the BSMI scales as $\log x$ where $x$ is defined in Eq.~\eqref{eq:x_scaling}. Moreover, the prefactors of these two logarithmic scalings are found to be approximately equal, giving another evidence of the conformal symmetry at the critical point. The fact that the BPMI exhibits similar logarithmic scaling behavior as the entanglement entropy suggests that, as the system undergoes an entanglement phase transition, there is a concurrent structural change in the SRE and PE, which is effectively captured by the behavior of the bipartite mutual information.

We also compute the BPMI defined in Eq.~\eqref{eq:mutual_pe}. Unlike the BSMI, the steady state BPMI seems to unexpectedly deviate from logarithmic behavior. This deviation appears to be a numerical artifact of the MPS simulation. Recalling that the bipartite mutual information is the difference between the entropy of the two subsystems and the whole system, we observe that the subsystems appear to be affected more severely by the bond dimension truncation. This effect is relatively small, i.e. $O(1)$, but it leads to a large correction in the mutual information. Unfortunately, a significantly larger bond dimension is not feasible in our simulation, so we relegate further analysis of the BPMI to the following section, in which we study a variation of the above model for purely Clifford dynamics.

\section{Clifford Circuits}\label{sec:clifford_circuit}

In previous sections, we investigated the critical behaviors of magic in hybrid circuits using a sampling approach in MPS. In this section, we turn to Clifford circuits, where we can simulate much larger system sizes. Since stabilizer states have exactly zero magic, we probe the critical behaviors using the participation entropy defined in Eq.~\ref{eq:pe}. We find that the PE exhibits similar critical slowing down as the magic.

The PE of a stabilizer state is easy to compute. First note that, for a stabilizer state, any nonzero measurement probability $p_z = \text{Tr}(\rho |z\rangle\langle z|)$ has the same value. In other words, the distribution in the computational basis is flat. Thus all R\'enyi participation entropies are the same, 
\begin{equation}
S_n^{\mathrm{PE}} = \log(\text{number of nonzero } p_z).
\end{equation}
The task reduces to counting how many basis states have nonzero probability.

The stabilizer states in the hybrid circuit, pure or mixed, all have the form $\rho = 2^{-n}\sum_{g\in G} g$, where $G$ is the stabilizer group and $n$ is the number of qubits in the (sub)system. The measurement probability is
\begin{equation}
p_z = 2^{-n}\sum_{g\in G} \langle z|g|z\rangle .
\end{equation}
Only $Z$-type Pauli strings contribute nonzero $\langle z|g|z\rangle$, so we restrict the sum to the subgroup $G_Z\subseteq G$. We have $p_z = 2^{-n}\langle z|\sum_{g\in G_Z} g|z\rangle$, and only those $|z\rangle$ that have $+1$ eigenvalues for all $g\in G_Z$ give nonzero probability. If there are $k_Z$ independent generators in $G_Z$, then there are $2^{n-k_Z}$ such $z$, and hence the participation entropy is $n-k_Z$.

The quantity $k_Z$ is the dimension of the subspace of Pauli-$Z$ stabilizers in $G$. It can be computed by linearly combining generators to eliminate their $X$ components, or equivalently, $k_Z = \dim \ker T_X$, where $T_X$ is the $X$-block of the stabilizer tableau. By the rank nullity theorem,
\begin{equation}
S_n^{\mathrm{PE}} = n - \dim\ker T_X = \text{rank}(T_X).
\end{equation}
For subsystem $A$, the tableau $T$ only includes part $A$ of the stabilizers. 

\subsection{Clifford circuit with duality symmetry}
\begin{figure}[h!]
  \centering
  \includegraphics[width=\linewidth]{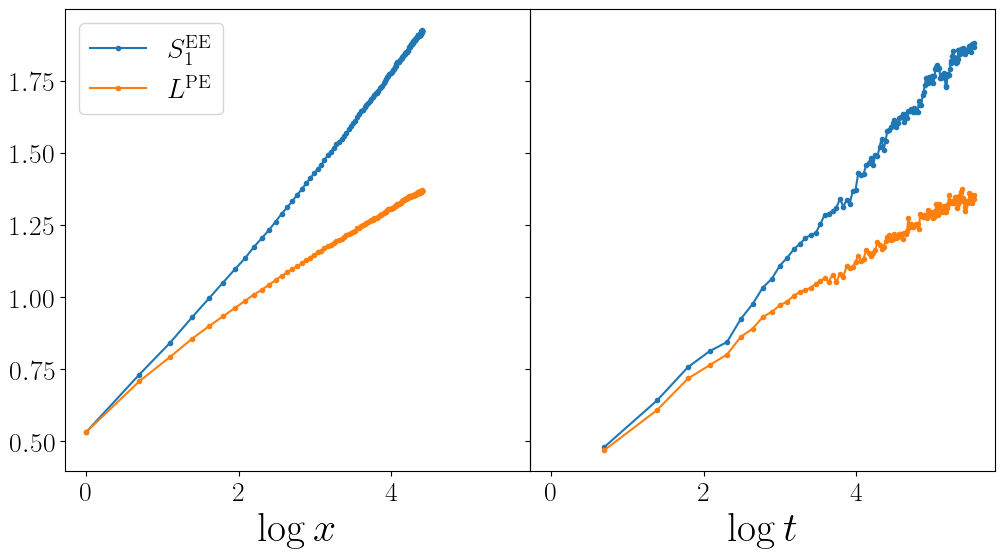}
  \caption{The growth of the entanglement and BPMI for the Clifford dynamics described in Sec.~\ref{sec:clifford_circuit} at $\gamma = 1$ and $L=256$, both displaying logarithmic growth in time and space. The scaling coefficients for entanglement are $\alpha_t^{\mathrm{EE}} = 0.31$, $\alpha_s^{\mathrm{EE}}=0.33$, yielding $z^\mathrm{EE}=1.06$. The scaling coefficients for the BPMI are $\alpha_t^\mathrm{PE} = 0.168$ and $\alpha_s^{\mathrm{PE}} = 0.167$, yielding $z^\mathrm{PE}=0.99$.
  }
  \label{fig:clifford_scaling}
\end{figure}

\begin{figure}[h!]
  \centering
  \includegraphics[width=\linewidth]{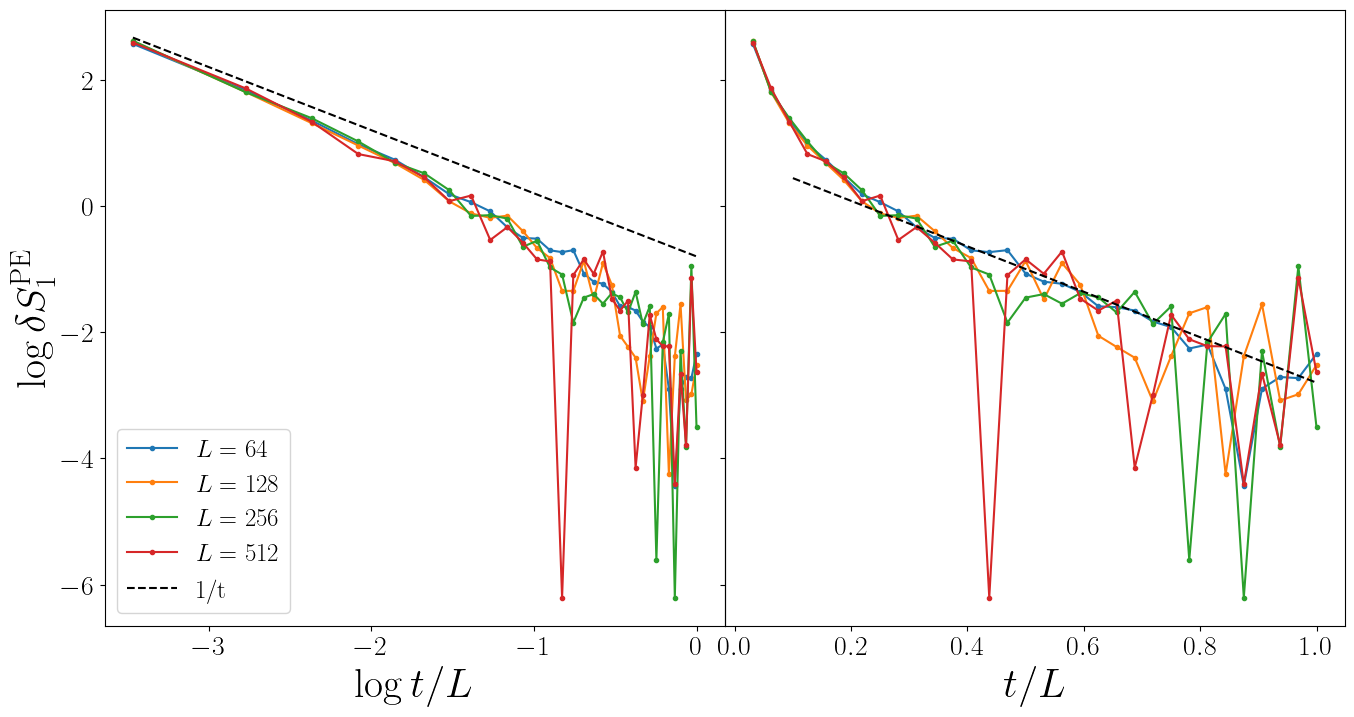}
  \caption{The convergence of the participation entropy of the Clifford model at $\gamma = 1$. Similarly to Fig.~\ref{fig:equilibrium}, we show the early-time evolution on the left, with a dotted guide line of slope 1 emphasizing the power-law scaling with unit slope, and the late-time exponential tail on the right. The guide-line has slope $\alpha=-3.6$.
  }
  \label{fig:clifford_equilibration}
\end{figure}

We study a Clifford variation of the self-dual hybrid circuits (Sec.~\ref{sec:self_dual})  by replacing the weak measurements of strength $\beta$ with projective measurements in the $XX$ or $Z$ basis with probability $\gamma$. 
In other words, at the measurement round, we randomly choose $L \gamma / 2$ pairs of neighboring sites and perform projective measurements. The $\gamma=1$ limit of this model is exactly the $\beta\rightarrow\infty$ limit of the non-Clifford model. Thus we can access the same type of the self-dual critical points in a Clifford model  through the participation entropy without the numerical challenges associated with MPS simulations. 
In this subsection, we investigate the critical line of this Clifford variant.

We first consider the critical point at $\gamma=1$ and $p=0.5$. Similarly to the non-Clifford case, we observe a slow relaxation to equilibrium PE, i.e. we find that 
\begin{equation}
\begin{aligned}
     \delta S_n^{\mathrm{PE}}(t)
  &= |S_n^{\mathrm{PE}}(t) - S_n^{\mathrm{PE}}(\infty)|\\
  &\;\propto\;
  \begin{cases}
    \tau^{-1}, & \tau \ll 1,\\[4pt]
    \exp(-\alpha\,\tau), & \tau \gg 1,
  \end{cases}
\end{aligned}
\end{equation}
see the numerical data in Fig.~\ref{fig:clifford_equilibration}. 

The BPMI at this critical point ($\gamma = 1$ and $p = 0.5$) exhibits the same logarithmic scaling as the entanglement in both time and space, as shown in Fig.~\ref{fig:clifford_scaling}. Such logarithmic growing behavior indicates that the PE develops non-local structure at criticality. Furthermore, the matching coefficients of the logarithmic growth in the temporal and spatial directions suggest that the Clifford model retains an emergent conformal symmetry at the critical point.
 
\begin{figure}[h!]
  \centering
  \includegraphics[width=\linewidth]{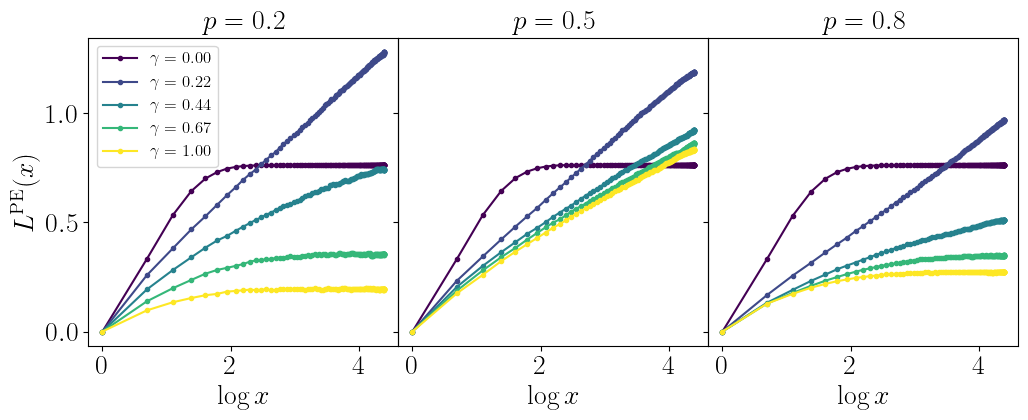}
  \caption{The BPMI of the Clifford circuit with duality symmetry at various model parameters. At $p=0.5$ the projective component and hence the full dynamics is self-dual, and the critical line is stable for any $\gamma \geq \gamma_c(p=0.5)$, as evidenced by the logarithmic scaling of the BPMI. Meanwhile, for $p \neq 0.5$, we see that the BPMI saturates to a finite constant in both the area-law and the volume-law, and only shows logarithmic scaling at $\gamma \approx 0.22$, which is roughly at the volume-law to area-law transition.}
  \label{fig:clifford_volume_pe}
\end{figure}

Finally, we vary the measurement probability $\gamma $  away from $1$ (See Fig.~\ref{fig:clifford_volume_pe}). In particular, we explore the entanglement phase transition from the area-law phase to the volume-law phase at fixed $p$. For the steady state, we find that at the critical point separating the entangled volume-law phase from the disentangled area-law phase, BPMI again displays logarithmic scaling similar to the entanglement entropy, whereas in both the area-law and volume-law phases, BPMI obeys only area-law scaling. 

\subsection{Random Clifford Circuit and Random Automaton Circuit}

We use the same formalism to study the PE of another two classes of one-dimensional hybrid Clifford circuits, each exhibiting an entanglement phase transition driven by projective measurements.

\begin{figure}[ht]
  \centering
  \includegraphics[width=0.5\textwidth]{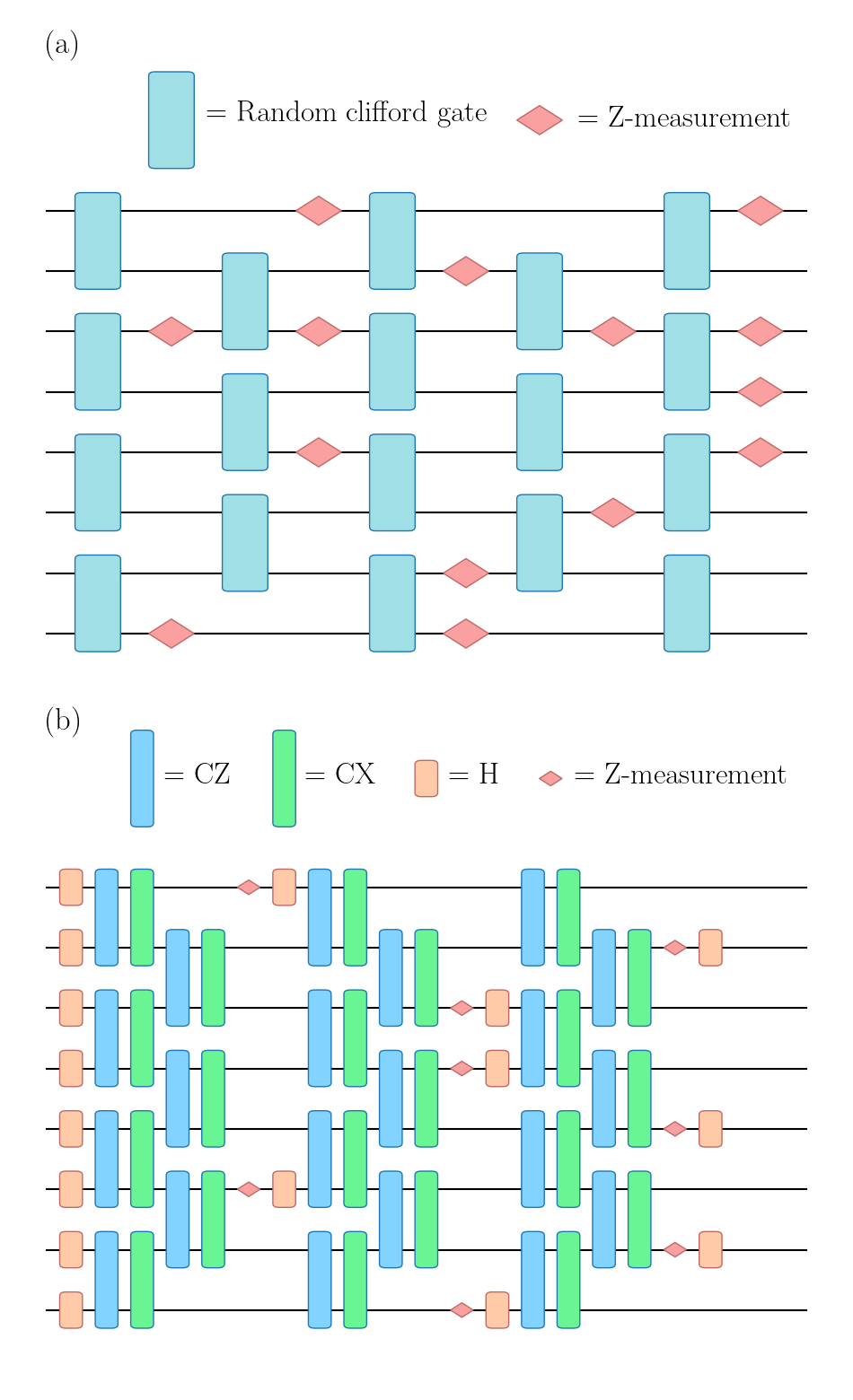}
  \caption{(a) A brickwork of two-qubit gates drawn uniformly at random from the Clifford group. After each layer, each qubit is measured in the $Z$-basis with probability $p$. The initial state is $\bigotimes\limits_i\ket{0}$.
    (b) A brickwork of QA gates, i.e. a $CZ$ gate followed by a $CX$ gate with the control and target qubits selected randomly with equal probability. After every two layers, every qubit is measured in the $Z$-basis immediately followed by a Hadamard gate with probability $p$. The initial state is $\bigotimes\limits_i \ket{+}$.}
    \label{fig:clifford_circuits}
\end{figure}

The first circuit (Fig.~\ref{fig:clifford_circuits}(a)) is the standard hybrid random Clifford circuit for MIPT~\cite{li2019_mipt}. The circuit consists of  two-qubit random Clifford gates in a brickwork pattern. After each layer of unitary gates, each qubit is measured in the $Z$-basis with probability $p$. This model exhibits a transition from a volume-law to an area-law phase.
\begin{equation}
    S_A \sim \left\{
    \begin{matrix}
        |A|       & p < p_c \approx 0.16,\\
        \log |A|  & p = p_c,\\
        \mathcal{O}(1) & p > p_c.
    \end{matrix}
    \right.
\end{equation}

At the critical measurement rate $p_c = 0.16$,
the PE $S_n^{\mathrm{PE}}(t)$ relaxes as
\begin{equation}
\begin{aligned}
     \delta S_n^{\mathrm{PE}}(t)
  &= |S_n^{\mathrm{PE}}(t) - S_n^{\mathrm{PE}}(\infty)|\\
  &\;\propto\;
  \begin{cases}
    \tau^{-1}, & \tau \ll 1,\\[4pt]
    \exp(-\alpha\,\tau), & \tau \gg 1,
  \end{cases}
\end{aligned}
\end{equation}
with $\tau = t/L$ and $\alpha \approx 3.6$. 

The BPMI of one contiguous half of the chain scales logarithmically 
\begin{equation}
  I_{A\bar A}^{\mathrm{PE}}(t)\sim a\,\log t,
  \quad a\approx0.31.
\end{equation}
These behaviors are qualitatively the same as the Clifford-variant of the self-dual hybrid circuit. 

The last circuit we study (Fig.~\ref{fig:clifford_circuits}(b)) is a hybrid random quantum automaton (QA) circuit~\cite{han2023_qa}, where the unitary gates are arranged in a brickwork fashion. Each unitary consists of a CNOT gate followed by a CZ gate, with the control qubit of the CNOT randomly chosen to be either the left or right qubit with equal probability. These unitaries map computational $Z$-basis states to other basis states up to a phase, and are referred to as automaton gates. In the measurement layer, each qubit is measured in the $Z$-basis with probability $p$, and every measurement is immediately followed by a Hadamard gate.

Under this hybrid protocol, the steady state entanglement exhibits a phase transition that belongs to the directed percolation universality class~\cite{iaconis_mipt_2020}. Furthermore, the steady state becomes an equal weight superposition of all allowed computational basis states, with nontrivial entanglement structure encoded in the relative phases between them. Therefore, the automaton gate does not change the PE in the $Z$ computational basis, and we define $S_n^{\mathrm{PE}}(t)$ via the distribution of measurement outcomes in the \(X\) basis,
\begin{equation}
  p_x(t) = \Tr\bigl(\rho(t)\ket{x}\bra{x}\bigr).
\end{equation}

\begin{figure}[ht]
    \centering
    \includegraphics[width=\linewidth]{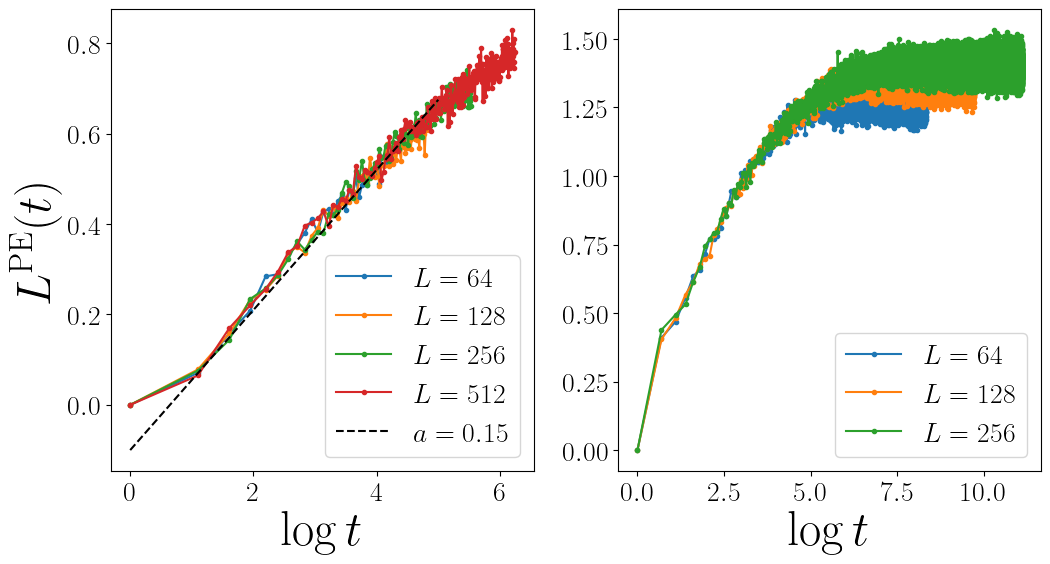}
    \caption{ Time evolution of the BPMI in (left) the hybrid Clifford circuit at $p_c=0.16$ and (right) the QA circuit at $p_c=0.138$.}
    \label{fig:rand_ca_time}
\end{figure}

In the QA circuit at its critical measurement rate $p_c = 0.138$, the relaxation of the participation entropy obeys
\begin{equation}
  \delta S_n^{\mathrm{PE}}(t)
  \;\propto\;
  \begin{cases}
    \tau^{-z},   & \tau \ll 1,\\[6pt]
    \exp(-\alpha \tau), & \tau \gg 1,
  \end{cases}
\end{equation}
where
$\tau = t/L^{\,z}$, $z = 1.64$, and $\alpha = 0.65$. 
We also study its BPMI at criticality. We find that BPMI (see Fig.~\ref{fig:rand_ca_time} (right)) also follows a strict logarithmic pattern, consistent with the other circuits.

\begin{figure}[ht]
    \centering
    \includegraphics[width=\linewidth]{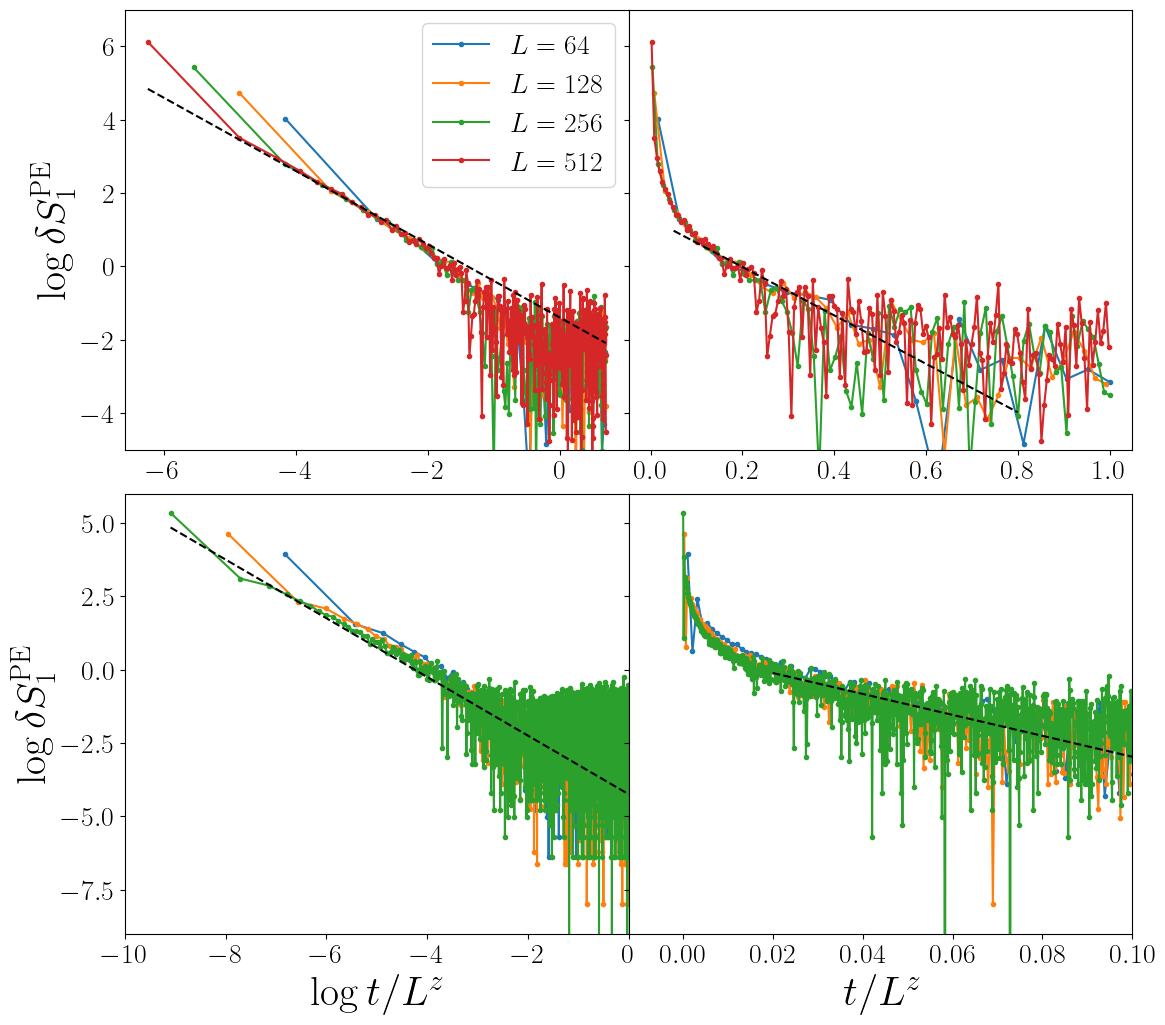}
    \caption{Top: PE relaxation in the random Clifford circuit at $p_c=0.16$. Bottom: PE relaxation in the QA circuit at $p_c=0.138$. }
    \label{fig:rand_ca_space}
\end{figure}

\section{Conclusion and Discussion}

In this work, we demonstrate that at the measurement induced critical points of non-unitary random circuits, both the PE and SRE  exhibit slow relaxation to equilibrium values. This behavior stands in sharp contrast to that of random unitary dynamics, where SRE and PE are known to approach equilibrium rapidly, typically within $O(\log L)$ time steps. We numerically confirm this slow relaxation in a quantum circuit possessing duality symmetry. In particular, we focus on the critical line separating two area-law phases, which allows simulations at comparatively large system sizes. We further compute the PE at critical points of several hybrid Clifford circuits and find the same scaling behavior.

Beyond the total SRE and PE, we also analyze their associated bipartite mutual information at criticality. We find that, analogous to entanglement entropy at measurement-induced critical points, the bipartite mutual information exhibits logarithmic scaling in both space and time, signaling the emergence of non-local correlations in both SRE and PE. Away from criticality, in both the volume-law and area-law phases, we show that the bipartite mutual information of PE obeys area-law scaling.

Our results indicate that measurement-induced entanglement transitions can be effectively characterized using other information theoretic quantities, such as PE and SRE. Beyond serving as diagnostics of MIPTs, these quantities also provide meaningful measures of the effective exploration of Hilbert space by quantum states. Looking ahead, it would be interesting to apply PE and SRE to a broader class of quantum phases, including symmetry broken phases, topologically ordered states, and spin glass phases, where nontrivial Hilbert-space structure and slow dynamics are expected to play an essential role.

\begin{acknowledgements}
   We gratefully acknowledge computing resources from Research Services at Boston College and the assistance provided by Wei Qiu.
    This research is supported by the National Science Foundation under Grant No. DMR-2219735 (H.L. and X.C.).
\end{acknowledgements}

\appendix
The appendices collect technical details used in the main text: Appendix A explains how duality pins the critical point, while Appendices B and C describe the MPS sampling algorithms used to evaluate SRE/SMI and PE/PMI.

\section{Pinning the Critical Point Under Duality}\label{app:majorana}

The Kramers-Wannier duality can be illustrated through a Jordan-Wigner transform, which maps the Pauli matrices to Majorana operators:
\begin{align}\label{eq:jordan_wigner}
    \gamma_{2i} &= K_i Y_i, \\
    \gamma_{2i + 1} &= K_i X_i,
\end{align}
where $K_i = \prod\limits_{j<i} Z_j$. These Majorana operators exist on a $1d$ lattice of size $2L$ and satisfy $\gamma_k^2=1$ and $\{\gamma_k, \gamma_\ell\} = 2\delta_{k\ell}$.
In this representation, Clifford dynamics maps Majoranas to Majoranas, the global $Z_2$ symmetry becomes Majorana-parity conservation, and the two measurement channels map to nearest-neighbor Majorana bilinears. The key identities are

\begin{align}
    \gamma_{2i -1} \gamma_{2i} &= Z_i, \\
    \gamma_{2i} \gamma_{2i+1} &= X_i X_{i+1}.
\end{align}

Thus, $Z_i$ and $X_iX_{i+1}$ measurements occupy even and odd Majorana bonds, respectively. A one-site translation in the Majorana chain exchanges these two bond types. At $p = 0.5$, the measurement ensemble is invariant under this translation, so the circuit is self-dual and the area-law-to-area-law critical point is pinned exactly at $p = 0.5$.

\section{Computing Stabilizer R\'enyi Entropy in Matrix Product States}\label{app:stabilizer_entropy}
This appendix gives an implementation-level summary of Pauli-string sampling from $\Pi_\rho(\boldsymbol{\sigma})$ in an MPS representation and the resulting estimators for SRE and SMI.

The Stabilizer R\'enyi Entropy is defined in Eq.~\eqref{eq:sre}. Note that $\Pi_\rho(\boldsymbol\sigma)$ defined in the main text forms a normalized probability distribution over $\mathcal{P}_L$.

Eq.~\eqref{eq:sre_est} is approximated by sampling Pauli strings from the $\Pi_\rho(\boldsymbol\sigma)$ distribution using the algorithm of Ref.~\cite{lami2023}, which we summarize here for completeness. The MPS is described by a right-normalized list of $L$ tensors $A_i^{s_i}$, where $s_i$ is the index corresponding to the qubit degree of freedom and the $A_i^{s_i}$ are $\chi \times \chi$ matrices for $ 1 < i < L$, while $A_1^{s_1}$ ($A_L^{s_L}$) is $1 \times \chi$ ($\chi \times 1$). These matrices are right normalized, i.e. 

\begin{align}
    \sum\limits_{s_i=0}^1 A_i^{s_i} (A_i^{s_i})^\dag = \mathbb{I}
\end{align}

Written diagrammatically,

\begin{align}
    \raisebox{-1.4cm}{
    \begin{tikzpicture}[tensor network]
        \node[triangle_r] (Ad) at (0,0) {$A_i^*$};
        \node[triangle_r] (A) at (0,2) {$A_i$};
        \begin{pgfonlayer}{background}
        \draw[edge]
            (A) -- (Ad)
            (A) -- ++(-0.9, 0)
            (Ad) -- ++(-0.9, 0)
            (Ad) -- ++(0.7, 0) coordinate (outAd)
            (A) -- ++(0.7, 0) coordinate (outA)
            (outAd) -- (outA);
        \end{pgfonlayer}
    \end{tikzpicture}
    } &=
    \raisebox{-0.9cm}{
    \begin{tikzpicture}[tensor network]
        \coordinate (outAd) at (0,0);
        \coordinate (outA) at (0,2);
        \begin{pgfonlayer}{background}
            \draw[edge] 
                (outAd) -- (outA)
                (outA) -- ++(-0.5, 0)
                (outAd) -- ++(-0.5, 0);
        \end{pgfonlayer}
    \end{tikzpicture}
    }
\end{align}

We write the distribution $\Pi_\rho(\boldsymbol{\sigma})$

\begin{align}\label{eq:m_est}
    \Pi_\rho(\boldsymbol{\sigma}) = \pi_\rho(\sigma_1) \pi_\rho(\sigma_2|\sigma_1) \cdots \pi_\rho(\sigma_L|\sigma_1 \cdots\sigma_{L-1})
\end{align}

The marginal distribution of the first site can be computed explicitly:

\begin{align}
    \pi_\rho(\sigma_1) = \frac{1}{2^N} \sum\limits_{\boldsymbol\sigma \in \mathcal{P}_{L-1}} \bra{\psi} \sigma_1\boldsymbol\sigma \ket{\psi} \bra{\psi^*} \sigma_1^* \boldsymbol\sigma^* \ket{\psi^*}
\end{align}

Using the right-orthonormality of $\ket{\psi}$ and the identity $\bra{s' r'} \sum\limits_{\sigma \in \mathcal{P}_1} \sigma \otimes \sigma^* \ket{s r} = \delta_{s' r'} \delta_{s r}$, this can be simplified and written in terms of the tensor network:

\begin{align}
\pi_\rho(\sigma_1) = \frac{1}{2} &
\raisebox{-1.4cm}{
  \begin{tikzpicture}[tensor network]
    \node[triangle_r] (Ad1) at (1,0) {$A_1^*$};
    \node[fill=red!20] (P1) at (1,1) {$\sigma_1$};
    \node[triangle_r] (A1) at (1,2) {$A_1$};
    \node[triangle_l] (Ad2) at (2,0) {$A_1$};
    \node[fill=red!20] (P2) at (2,1) {$\sigma_1^*$};
    \node[triangle_l] (A2) at (2,2) {$A_1^*$};
    \node[shape=rectangle, minimum height=8em, fill=green!20] (L1) at (0,1) {$L$};
    \node[shape=rectangle, minimum height=8em, fill=green!20] (L2) at (3,1) {$L^*$};
    \begin{pgfonlayer}{background}
    \draw[edge]
    (A1) -- (L1 |-  A1)
    (Ad1) -- (L1 |- Ad1)
    (A1) -- (P1)
    (Ad1) -- (P1)
    (A2) -- (P2)
    (Ad2) -- (P2)
    (Ad1) -- (Ad2)
    (A1) -- (A2)
    (A2) -- (L2 |-  A2)
    (Ad2) -- (L2 |- Ad2);
    \end{pgfonlayer}
  \end{tikzpicture}
} 
\end{align}

where we have additionally defined the "environment tensor" $L = (1)$. At the boundary, this
addition is redundant, as the boundary index has dimension 1. After evaluating $\pi_\rho(\sigma_1)$ for each of $\sigma_1=\sigma_I,\sigma_X,\sigma_Y,\sigma_Z$, we can draw a sample from this distribution and fix the first element of the Pauli string. We then partially project the state onto this choice, allowing us to continue drawing samples from each qubit in the string from the appropriate conditional distribution. This is accomplished by updating the environment tensor:

\begin{align}
    \raisebox{-1.5cm} {
    \begin{tikzpicture}[tensor network]
        \node[rectangle, minimum height=8em, fill=green!20] (L) at (0,1) {$L$};
        \begin{pgfonlayer}{background}
        \draw[edge] 
            (L)+(0,1) -- ++(1,1)
            (L)+(0,-1) -- ++(1,-1);
        \end{pgfonlayer}
    \end{tikzpicture} }&
    \mathrel{\rightarrow}
    \frac{1}{\sqrt{\pi_\rho(\sigma_1)}} \frac{1}{\sqrt{2}}
    \raisebox{-1.5cm} {
    \begin{tikzpicture}[tensor network]
        \node[triangle_r] (Ad) at (1,0) {$A_1^*$};
        \node[fill=red!20] (P) at (1,1) {$\sigma_1$};
        \node[triangle_r] (A) at (1,2) {$A_1$};
        \node[rectangle, minimum height=8em, fill=green!20] (L) at (0,1) {$L$};
        \begin{pgfonlayer}{background}
        \draw[edge]
            (A) -- (L |- A)
            (Ad) -- (L |- Ad)
            (A) -- (P)
            (Ad) -- (P)
            (A) -- ++(1,0)
            (Ad) -- ++(1,0);
        \end{pgfonlayer}
    \end{tikzpicture}
    }
\end{align}

The following Paulis are fixed by the same method, updating the environment tensor to account for the partial projection onto the previous choices for single-qubit Paulis. That is, the conditional distribution

\begin{align}\label{eq:sre_conditional_dist}
    \pi_\rho(\sigma_i | \sigma_1 \cdots \sigma_{i-1}) = \frac{1}{2} &
\raisebox{-1.4cm}{
  \begin{tikzpicture}[tensor network]
    \node[triangle_r] (Ad1) at (1,0) {$A_i^*$};
    \node[fill=red!20] (P1) at (1,1) {$\sigma_i$};
    \node[triangle_r] (A1) at (1,2) {$A_i$};
    \node[triangle_l] (Ad2) at (2,0) {$A_i$};
    \node[fill=red!20] (P2) at (2,1) {$\sigma_i^*$};
    \node[triangle_l] (A2) at (2,2) {$A_i^*$};
    \node[shape=rectangle, minimum height=8em, fill=green!20] (L1) at (0,1) {$L$};
    \node[shape=rectangle, minimum height=8em, fill=green!20] (L2) at (3,1) {$L^*$};
    \begin{pgfonlayer}{background}
    \draw[edge]
    (A1) -- (L1 |-  A1)
    (Ad1) -- (L1 |- Ad1)
    (A1) -- (P1)
    (Ad1) -- (P1)
    (A2) -- (P2)
    (Ad2) -- (P2)
    (Ad1) -- (Ad2)
    (A1) -- (A2)
    (A2) -- (L2 |-  A2)
    (Ad2) -- (L2 |- Ad2);
    \end{pgfonlayer}
  \end{tikzpicture}
}
\end{align}
is calculated for $\sigma_i = \sigma_I,\sigma_X,\sigma_Y,\sigma_Z$. Then, a sample from the single-site distribution is chosen, and the environment tensor is updated accordingly:
\begin{align}\label{eq:sre_env_update}
    \raisebox{-1.5cm} {
    \begin{tikzpicture}[tensor network]
        \node[rectangle, minimum height=8em, fill=green!20] (L) at (0,1) {$L$};
        \begin{pgfonlayer}{background}
        \draw[edge] 
            (L)+(0,1) -- ++(1,1)
            (L)+(0,-1) -- ++(1,-1);
        \end{pgfonlayer}
    \end{tikzpicture} }&
    \mathrel{\rightarrow}
    \frac{1}{\sqrt{\pi_\rho(\sigma_i | \sigma_1 \cdots \sigma_{i - 1})}} \frac{1}{\sqrt{2}}
    \raisebox{-1.5cm} {
    \begin{tikzpicture}[tensor network]
        \node[triangle_r] (Ad) at (1,0) {$A_i^*$};
        \node[fill=red!20] (P) at (1,1) {$\sigma_i$};
        \node[triangle_r] (A) at (1,2) {$A_i$};
        \node[rectangle, minimum height=8em, fill=green!20] (L) at (0,1) {$L$};
        \begin{pgfonlayer}{background}
        \draw[edge]
            (A) -- (L |- A)
            (Ad) -- (L |- Ad)
            (A) -- (P)
            (Ad) -- (P)
            (A) -- ++(1,0)
            (Ad) -- ++(1,0);
        \end{pgfonlayer}
    \end{tikzpicture}
    }
\end{align}

The algorithm is written in pseudocode in Alg.~\ref{alg:sre_mps}. The resulting Pauli strings are distributed without bias according to $\Pi_\rho(\boldsymbol\sigma)$, and therefore the SRE can be estimated using the right side of Eq.~\eqref{eq:m_est}. Since each iteration requires $O(L)$ multiplications of $\chi \times \chi$ matrices, the overall runtime of the algorithm is $O(L\chi^3)$ per Pauli string sampled, and estimating the SRE is $O(N_sL\chi^3)$, where $N_s$ is the number of Pauli strings sampled.

\begin{figure}[t]
\caption{Pauli string sampling of MPS}\label{alg:sre_mps}
\begin{algorithmic}[1]
\Function{SamplePauliString}{$\ket{\psi}$}
\State Put $\ket{\psi}$ in right-orthogonalized form
\State $\Pi \gets 1$
\State $L \gets (1)$
\For {i = 0; i < L; i++}
    \State Compute $\pi(\alpha)=\pi_\rho(\sigma_\alpha|\sigma_1\cdots\sigma_{i-1})$ as shown in Eq.~\ref{eq:sre_conditional_dist}.
    \State Fix $\sigma_i = \sigma_\alpha$ with probability $\pi(\alpha)$.
    \State Update $L$ as shown in Eq.~\ref{eq:sre_env_update}.
    \State Update $\Pi \rightarrow \Pi\cdot\pi(\alpha)$
\EndFor
\State \Return{$\Pi,\sigma_1\cdots\sigma_L$}
\EndFunction
\end{algorithmic}
\end{figure}

We compute the SRE and SMI using Paulis sampled from $\Pi_\rho$. The SMI, defined in Eq.~\eqref{eq:mutual_sre} for $n = 2$ on subsystems $A$ and $B$ can be written as 

\begin{align*}
    L^{\mathrm{SRE}}_{AB}(\rho) &= S_2^{\mathrm{SRE}}(\rho_A) + S_2^{\mathrm{SRE}}(\rho_B) - S_2^{\mathrm{SRE}}(\rho_{AB}) \\
    &= I(\rho_{AB}) - W(\rho_{AB})
\end{align*}
where
\begin{align*}
    I(\rho_{AB}) &= -\log\left\langle \frac{|\text{Tr}(\rho_A \boldsymbol\sigma_A)|^2 |\text{Tr}(\rho_B\boldsymbol\sigma_B)|^2} {|\text{Tr}(\rho_{AB}\boldsymbol\sigma_{AB})|^2} \right\rangle_{\Pi_\rho} \\
    W(\rho_{AB}) &= -\log\left\langle \frac{|\text{Tr}(\rho_A \boldsymbol\sigma_A)|^4 |\text{Tr}(\rho_B \boldsymbol\sigma_B)|^4} {|\text{Tr}(\rho_{AB}\boldsymbol\sigma_{AB})|^4} \right\rangle_{\Xi_\rho}
\end{align*}
where $\boldsymbol\sigma_A$ ($\boldsymbol \sigma_B$) are the restrictions of $\boldsymbol\sigma_{AB}$ onto the subsystem $A$ ($B$), and $\Xi_\rho(\boldsymbol\sigma) \sim |\text{Tr}(\rho \boldsymbol\sigma)|^4$. Since the above algorithm can only directly sample from $\Pi_\rho$, we slightly modify the above expression by re-weighting the sum in the $W(\rho_{AB})$ term and arrive at
\begin{align*}
    L^{\mathrm{SRE}}_{AB}(\rho) &= I(\rho_{AB}) - \widetilde{W}(\rho_{AB}) + S^{\mathrm{SRE}}_2(\rho_{AB})
\end{align*}
where
\begin{align*}
    \widetilde{W}(\rho_{AB}) &= -\log\left\langle \frac{|\text{Tr}(\rho_A \boldsymbol\sigma_A)|^4 |\text{Tr}(\rho_B\boldsymbol\sigma_B)|^4} {|\text{Tr}(\rho_{AB}\boldsymbol\sigma_{AB})|^2} \right\rangle_{\Pi_\rho}. \\
\end{align*}

\section{Computing Participation Entropy in Matrix Product States}\label{app:participation_entropy}
This appendix presents the analogous bitstring-sampling procedure in the computational basis, then uses those samples to estimate PE and PMI.

The participation entropy is defined in Eq.~\eqref{eq:pe}. The probability amplitudes $p(z) = |\bra{z}\ket{\psi}|^2$ of an MPS state $\ket{\psi}$ is estimated by directly sampling bitstrings and their corresponding probability amplitudes. 
\begin{align}\label{eq:w_est}
    \widetilde{S}^{\mathrm{PE}}(\rho) = -\sum\limits_{i=0}^{N-1} \log_2 p(z^{(i)})
\end{align}
where $z^{(i)}$ are sampled from the underlying MPS. The bitstring sampling is accomplished using a technique very similar to the technique for sampling Pauli strings in the previous section, whereby a bitstring is constructed bit-by-bit such that at site $i$, the bitstring constructed thus far is sampled from the marginal distribution $p(z_1 \cdots z_i)$, where $z_i$ is the $i$th bit of the integer $z$, and thus $z$ can be written as the string $z_1 \cdots z_L$. We assume that the state is an MPS in the right-normalized form as in the previous section. The first site can be easily sampled by calculating $p(z_1 = \alpha) = \langle P_\alpha\rangle$, where $P_\alpha = \ket{\alpha} \bra{\alpha}$ acting on the first qubit. Using right-normalization, this requires contracting only the leftmost tensor with the projector, i.e. by computing the contraction 
\begin{align}
    p(z_1 = \alpha) = &
    \raisebox{-1.5cm} {
    \begin{tikzpicture}[tensor network]
        \node[rectangle, minimum height=8em, fill=green!20] (L) at (0,1) {$L$};
        \node[triangle_r] (Ad) at (1,0) {$A_1^*$};
        \node[fill=red!20] (P) at (1,1) {$P_\alpha$};
        \node[triangle_r] (A) at (1,2) {$A_1$};
        \begin{pgfonlayer}{background}
        \draw[edge]
            (A) -- (L |- A)
            (Ad) -- (L |- Ad)
            (Ad) -- (P)
            (A) -- (P)
            (Ad) -- ++(0.7,0) coordinate (outAd)
            (A) -- ++(0.7,0) coordinate (outA);
        \draw[edge]
            (outAd) -- (outA);
        \end{pgfonlayer}
    \end{tikzpicture}
    }
\end{align}
where we have defined the "environment tensor" $L = (1)$. Similarly to above, this environment tensor is redundant for the first iteration, but will be updated at later iterations. After $p(z_1=0)$ and $p(z_1=1)$ are calculated, $z_1$ is fixed to $z_\alpha$ with probability $p(z_1=z_\alpha)$. We then update $L$ according to

\begin{align}
    \raisebox{-1.5cm} {
    \begin{tikzpicture}[tensor network]
        \node[rectangle, minimum height=8em, fill=green!20] (L) at (0,1) {$L$};
        \begin{pgfonlayer}{background}
        \draw[edge] 
            (L)+(0,1) -- ++(1,1)
            (L)+(0,-1) -- ++(1,-1);
        \end{pgfonlayer}
    \end{tikzpicture} }&
    \mathrel{\rightarrow}
    \frac{1}{p(z_1)}
    \raisebox{-1.5cm} {
    \begin{tikzpicture}[tensor network]
        \node[triangle_r] (Ad) at (1,0) {$A_1^*$};
        \node[fill=red!20] (P) at (1,1) {$P_{z_1}$};
        \node[triangle_r] (A) at (1,2) {$A_1$};
        \node[rectangle, minimum height=8em, fill=green!20] (L) at (0,1) {$L$};
        \begin{pgfonlayer}{background}
        \draw[edge]
            (A) -- (L |- A)
            (Ad) -- (L |- Ad)
            (A) -- (P)
            (Ad) -- (P)
            (A) -- ++(1,0)
            (Ad) -- ++(1,0);
        \end{pgfonlayer}
    \end{tikzpicture}
    }
\end{align}

The subsequent sites are sampled using the recursion relation relating the marginal distributions of adjacent sites:

\begin{align}\label{eq:recursive_marginal}
    p(z_1 \cdots z_i) = p(z_i | z_1 \cdots z_{i-1}) p(z_1 \cdots z_{i-1})
\end{align}

We can compute $p(z_i|z_1 \cdots z_{i-1})$ using the results from the previous iteration as follows. Suppose $L$ represents the partial projection of the tensor onto the bits measured thus far. This tensor corresponds to the MPS projected onto the $z_1 \cdots z_{i-1}$ state and partially contracted with itself. Therefore, contracting $L$ with $P_\alpha$ on the $i$th site gives the conditional probability of measuring $z_i$. That is,

\begin{align}\label{eq:config_mps}
    p(z_i = \alpha | z_1 \cdots z_{i - 1}) = &
    \raisebox{-1.5cm} {
    \begin{tikzpicture}[tensor network]
        \node[rectangle, minimum height=8em, fill=green!20] (L) at (0,1) {$L$};
        \node[triangle_r] (Ad) at (1,0) {$A_i^*$};
        \node[fill=red!20] (P) at (1,1) {$P_\alpha$};
        \node[triangle_r] (A) at (1,2) {$A_i$};
        \begin{pgfonlayer}{background}
        \draw[edge]
            (A) -- (L |- A)
            (Ad) -- (L |- Ad)
            (Ad) -- (P)
            (A) -- (P)
            (Ad) -- ++(0.7,0) coordinate (outAd)
            (A) -- ++(0.7,0) coordinate (outA);
        \draw[edge]
            (outAd) -- (outA);
        \end{pgfonlayer}
    \end{tikzpicture}
    }
\end{align}

Using the conditional probabilities, we sample $z_i$, calculate the updated marginal $p(z_1 \cdots z_i)$ using Eq.~\ref{eq:recursive_marginal}, and continue to the next iteration, updating the environment tensor:
\begin{align}\label{eq:config_env_update}
    \raisebox{-1.5cm} {
    \begin{tikzpicture}[tensor network]
        \node[rectangle, minimum height=8em, fill=green!20] (L) at (0,1) {$L$};
        \begin{pgfonlayer}{background}
        \draw[edge] 
            (L)+(0,1) -- ++(1,1)
            (L)+(0,-1) -- ++(1,-1);
        \end{pgfonlayer}
    \end{tikzpicture} }&
    \mathrel{\rightarrow}
    \frac{1}{p(z_1 \cdots z_i)}
    \raisebox{-1.5cm} {
    \begin{tikzpicture}[tensor network]
        \node[triangle_r] (Ad) at (1,0) {$A_1^*$};
        \node[fill=red!20] (P) at (1,1) {$P_{z_1}$};
        \node[triangle_r] (A) at (1,2) {$A_1$};
        \node[rectangle, minimum height=8em, fill=green!20] (L) at (0,1) {$L$};
        \begin{pgfonlayer}{background}
        \draw[edge]
            (A) -- (L |- A)
            (Ad) -- (L |- Ad)
            (A) -- (P)
            (Ad) -- (P)
            (A) -- ++(1,0)
            (Ad) -- ++(1,0);
        \end{pgfonlayer}
    \end{tikzpicture}
    }
\end{align}

Once every bit has been fixed, we return the bitstring and the computed value of $p(z_1 \cdots z_L)$. These sampled probabilities are then passed to Eq.~\eqref{eq:w_est} to estimate the configurational entropy. The bitstring sampling algorithm is described in pseudocode by Alg.~\ref{alg:ce_mps}. Since each interaction requires $O(L)$ multiplications of $\chi\times\chi$ matrices, the total runtime is $O(L\chi^3)$ per bitstring sampled.  

\begin{figure}[t]
\caption{Bitstring sampling of MPS}\label{alg:ce_mps}
\begin{algorithmic}[1]
\Function{SampleBitstring}{$\ket{\psi}$}
\State Put $\ket{\psi}$ in right-orthogonalized form
\State $p \gets 1$
\State $L \gets (1)$
\For {i = 0; i < L; i++}
    \State Compute $p(\alpha) = p(z_\alpha | z_{i-1}, z_{i-2}, ...)$ as shown in Eq.~\ref{eq:config_mps}.
    \State Fix $z_i = z_\alpha$ with probability $p(\alpha)$.
    \State Update $L$ with $P_{z_\alpha}$ as shown in Eq.~\ref{eq:config_env_update}.
    \State $p \gets p \cdot p(\alpha)$
\EndFor
\State \Return{$p$}
\EndFunction
\end{algorithmic}
\end{figure}

The PE and PMI are computed using samples from the above algorithm. The PE is calculated using Eq.~\ref{eq:w_est}. The PMI for subsystems $A$ and $B$ is defined in Eq.~\eqref{eq:mutual_pe}, and can be written

\begin{align}
    L_{AB}^{\mathrm{PE}}(\rho) &= S^{\mathrm{PE}}(\rho_A) + S^{\mathrm{PE}}(\rho_B) - S^{\mathrm{PE}}(\rho_{AB}) \\
    &= -\sum\limits_{z_A z_B} p(z_A z_B) \log_2 \left[ \frac{p(z_A) p(z_B)}{p(z_A z_B)} \right] \\
    &= \left\langle \log_2 \left[ \frac{p(z_A z_B)}{p(z_A) p(z_B)} \right] \right\rangle_{p(z_A z_B)}
\end{align}
where $z_A$ ($z_B$) are the restrictions of $z_{AB}$ onto $A$ ($B$), and $z_A z_B$ is understood as the concatenation of the two bitstrings.

\bibliography{PE_SRE.bib}

\end{document}